\documentclass[a4paper,11pt]{article}
\pdfoutput=1
\usepackage{jheppub}
\usepackage[usenames,dvipsnames]{xcolor}
\usepackage{slashed}
\usepackage[T1]{fontenc}
\usepackage[ansinew]{inputenc}
\usepackage{amsmath}
\usepackage{amssymb}
\usepackage{graphicx}
\usepackage{color}
\definecolor{darkblue}{cmyk}{0.9,0.9,0,0}
\definecolor{darkgreen}{rgb}{0,0.55,0}
\usepackage{minitoc}
\usepackage[normalem]{ulem}
\usepackage{enumitem}
\usepackage{tikz-cd}

\makeatletter

\newcommand{\comment}[1]{}

\newcommand{\beq}{\begin{equation}}
\newcommand{\eeq}{\end{equation}}
\newcommand{\beqq}{\begin{equation*}}
\newcommand{\eeqq}{\end{equation*}}
\newcommand\beqa{\begin{eqnarray}}
\newcommand\eeqa{\end{eqnarray}}
\newcommand\beqaa{\begin{eqnarray*}}
	\newcommand\eeqaa{\end{eqnarray*}}
\newcommand\bea{\begin{array}}
	\newcommand\eea{\end{array}}

\def\XXint#1#2#3{{\setbox0=\hbox{$#1{#2#3}{\int}$ }
		\vcenter{\hbox{$#2#3$ }}\kern-.5\wd0}}

\def\XXint#1#2#3{{\setbox0=\hbox{$#1{#2#3}{\int}$}
		\vcenter{\hbox{$#2#3$}}\kern-.5\wd0}}

\newcommand{\nn}{\nonumber}

\newcommand{\neqa}{\nonumber\end{eqnarray}}
\newcommand{\la}[1]{\label{#1}}

\def\tr{{\rm tr~}}

\newcommand{\hs}{\frac{\sqrt{3}}{2}}
\renewcommand{\d}{\partial}

\newcommand{\<}{{\langle}}
\renewcommand{\>}{{\rangle}}

\newcommand{\cC}{{\cal C}}
\newcommand{\cD}{{\cal D}}
\newcommand{\cL}{{\cal L}}

\newcommand{\re}{\relax{\rm I\kern-.18em R}}

\renewcommand{\sp}{p\hspace{-.40em}/}

\def\su2{{SU(2)}}

\def\eps{{\epsilon}}

\def\a{{\alpha}}

\def\[{\left[}
\def\]{\right]}

\def\l{\lambda}
\def\e{\epsilon}

\def\s{\sigma}
\def\a{\alpha}
\def\b{\Bethe}
\def\th{\theta}

\def\l{\lambda}
\def\e{\epsilon}

\def\s{\sigma}
\def\a{\alpha}
\def\b{\beta}
\def\th{\theta}
\def\D{\Delta}
\def\g{\gamma}

\def\({\left(}
\def\){\right)}
\def\[{\left[}
\def\]{\right]}

\def\<{\langle}
\def\>{\rangle}

\def\cO{{\cal O}}

\def\cC{{\cal C}}
\def\cW{{\cal W}}
\def\cP{{\cal P}}
\def\cM{{\cal M}}

\def\s*{\ *_{\!\!\!\!\!\!\!\!\!\,_{\,_\text{\scriptsize{sym}}}}}
\def\hs*{\ \hat{*}_{\!\!\!\!\!\!\!\!\!\,_{\,_\text{\scriptsize{sym}}}}}
\def\d{\partial}

\def\i2{\frac{i}{2}}

\def\spi{\relax{\rm \pi\kern-0.5em /}}
\def\sA{\relax{\rm A\kern-0.5em /}}
\def\sp{\relax{\rm p\kern-0.5em /}}
\def\sd{\relax{\rm \d\kern-0.5em /}}
\def\sk{\relax{\rm k\kern-0.5em /}}
\def\sn{\relax{\rm n\kern-0.5em /}}
\def\sl{\relax{\rm l\kern-0.5em /}}
\def\sP{\relax{\rm P\kern-0.7em /}}
\def\sBethe{\relax{\rm \Bethe\kern-0.5em /}}

\def\be#1\ee{\begin{equation}\begin{aligned}
#1
\end{aligned}
\end{equation}}

\newcommand{\ii}{\mathrm{i}}
\newcommand{\dd}{\mathrm{d}}

\newcommand{\llangle}{\<\!\<}
\newcommand{\rrangle}{\>\!\>}

\numberwithin{equation}{section}
\usepackage{varioref}
\usepackage{makeidx}
\usepackage{cleveref}
\makeindex

\title{Planar RG Flows on Line Defects}

\author[a]{Ivri Nagar,}
\author[a]{Amit Sever,}
\author[b]{and De-liang Zhong}
\affiliation[a]{School of Physics and Astronomy, Tel Aviv University, Ramat Aviv 69978, Israel}
\affiliation[b]{Blackett Laboratory, Imperial College, Prince Consort Road, London, SW7 2AZ, UK}

\abstract{
We study a class of renormalization group flows on line defects that can be described by a generalized free field with ordered planar contractions on the line. They are realized, for example, in large $N$ gauge theories with matter in the fundamental representation and arise generically in non-relativistic CFTs. We analyze the flow exactly and compute the change in the $g$-function between the UV and IR fixed points. We relate the result to the change in the two-point function of the displacement operator and check the monotonicity of the defect entropy along the flow analytically. Finally, we give a general realization of this type of flow starting from the direct sum of the IR fixed point and a trivial line. This type of defect renormalization group flow parallels the well-studied case of double-trace flow.}

\begin{document} 
\maketitle
\flushbottom

\section{Introduction}

We study renormalization group (RG) flows that take place along a line defect in large $N$ conformal field theories (CFTs) in $d\ge2$ dimensions. We focus on straight lines or circular defects. At a fixed point of these flows, the defect preserves an $SL(2,{\mathbb R})\times SO(d-1)$ subgroup of the $d$ dimensional conformal symmetry. The combined system is called a defect CFT (DCFT). The defect RG flow can be triggered by deforming the action with a local relevant defect operator ${\mathbb O}(x)$ as
\beq\la{deltaS}
S\ \rightarrow\ 
S+\lambda\,M^{1-\Delta_{\mathbb O}}\int \dd x\,{\mathbb O}(x)\,.
\eeq
Here, $\lambda$ is the deformation parameter, $M$ is the mass scale of the flow, and $\Delta_{\mathbb O}$ is the dimension of the operator at $\lambda=0$. The operator is relevant for $\Delta_{\mathbb O}<1$. The RG flow can take place close to the defect without affecting the CFT far from it. In this case, it must end at a new fixed point. 
The flow can also destabilize the line and affect the infrared. 

\begin{figure}[t]
    \centering
    \includegraphics[width=\textwidth ]{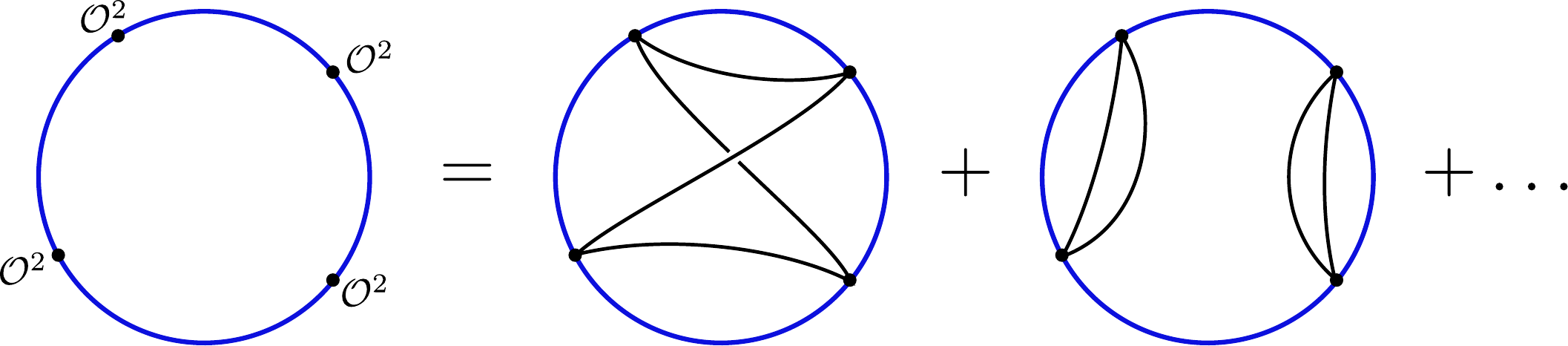}
    \caption{Four-point correlation function of the double-trace operator on the circle. At leading order in the large $N$ limit it factorizes into a product of the two-point functions of the single-trace operators. All possible contractions of the single-trace operators contribute.}
    \label{abelianfig}
\end{figure}

There is already extensive work on defect RG (DRG) flows; see \cite{Billo:2016cpy, SoderbergRousu:2023ucv,Cuomo:2021rkm,Affleck:1991tk, Dorey:1999cj,Yamaguchi:2002pa,Friedan:2003yc,Azeyanagi:2007qj,Takayanagi:2011zk,Estes:2014hka,Gaiotto:2014gha,Jensen:2015swa,Casini:2016fgb,Andrei:2018die,Kobayashi:2018lil,Casini:2018nym,Giombi:2020rmc,Wang:2020xkc,Nishioka:2021uef,Wang:2021mdq,Sato:2021eqo}. DRG flows that occur in large $N$ CFTs are simpler to analyze due to their simplified diagrammatics. 
One class of such simple DRGs is triggered by a double-trace operator
\beq\la{dtdeformation}
{\mathbb O}_\text{DT}(x)=\cO(x)^2+O(1/N)\,.
\eeq
Here, $\cO$ is a single-trace operator. 
In this case $\Delta_{\mathbb O}=2\Delta+O(1/N)$, where $\Delta$ is the dimension of $\cO$. What makes this flow particularly simple is the large $N$ factorization of the correlators of the double-trace operator into products of two-point functions of its single-trace constituents; see figure \ref{abelianfig}. Double-trace flow was studied in \cite{Giombi:2018vtc, Gubser:2002vv, Hartman:2006dy, Diaz:2007an}, and the holographic dual of this flow was studied in \cite{Witten:2001ua, Berkooz:2002ug, Hartman:2006dy,  Diaz:2007an, Giombi:2013yva,Giombi:2018vtc}. They can be realized on defects of any dimension, not necessarily a line. The flow triggered by (\ref{dtdeformation}) depends only on $\Delta$ and on the dimension of the defect, but not on the space-time dimension of the theory in which it is embedded; therefore, the results of these works apply equally well to defects.

Our focus in this paper is on a second class of DRG flows. These are flows that appear in the large $N$ limit of conformal gauge theories with matter in the fundamental representation of the gauge group. In such theories, a conformal defect in the fundamental representation can be deformed by an operator of the form 
\beq\la{FantiF}
{\mathbb O}_{\bar FF}=\overline{\cO}(x) \times \cO(x)\,,
\eeq
where $\cO$ is an operator that transforms in the fundamental representation of the gauge group and $\overline\cO$ in the anti-fundamental representation.\footnote{Many of the results of this paper can be generalized to deformations by defect-changing operators \cite{Billo:2013jda}. We leave these for future work.} The operator ${\mathbb O}_{\bar FF}$ is in the adjoint representation and exists only on the defect.

\begin{figure}[!ht]
    \centering
    \includegraphics[width=0.62\textwidth]{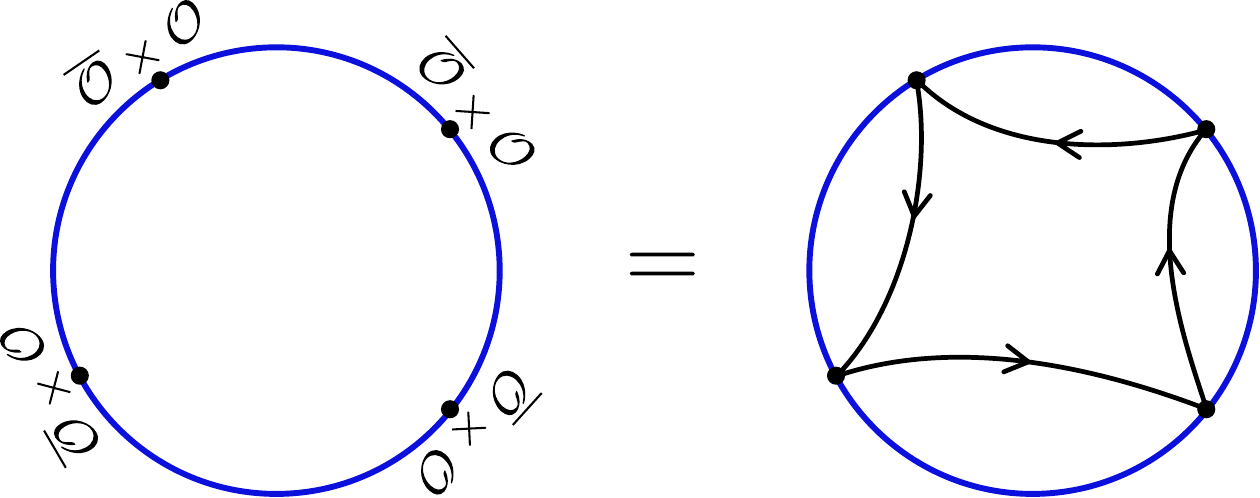}
    \caption{Four-point correlation function of the fundamental--anti-fundamental operator on the circle. At leading order in the large $N$ limit it factorizes into a product of the two-point functions between the fundamental and anti-fundamental operators. Only connected planar contractions contribute, joining together into a single fundamental loop.}
    \label{unregulated}
\end{figure}

As for the double-trace deformation, $\Delta_{\mathbb O}=2\Delta+O(1/N)$, and the correlators of the operator in (\ref{FantiF}) factorize into a product of two-point functions of its constituents at large $N$. However, there are two key differences from the double-trace case (\ref{dtdeformation}). 
First, only products of two-point functions that are both planar and connected survive the large $N$ limit (the products must be connected because additional disconnected components form additional fundamental loops).
These are two-point functions between neighboring ordered points on the line, see figure \ref{unregulated}. Second, the planar correlators of the operator (\ref{FantiF}) are of order one, while the expectation value of the undeformed line is of order $N$. This is because they form a fundamental loop.

Another class of examples where such a DRG flow is realized is non-relativistic theories. In these cases, the operator $\overline\cO$ annihilates a particle, while the operator $\cO$ creates one. The planar factorization of the correlators of ${\mathbb O}_{\bar FF}$ on a timelike defect follows from the conservation of the particle number in the non-relativistic limit. This class of DRG flows also extends to higher-dimensional defects.

One characteristic of a defect is the defect entropy $s$. For relativistic theories it is given by\footnote{An alternative definition of the defect entropy, that also extends to non-relativistic CFTs, is as (the logarithm of) the ratio of the partition functions of the theory on $S^{d-1}\times{\mathbb R}_t$, with and without two conjugate defects at the south and north poles.} 
\beq
s(RM) = (1-R\partial_R)\log\<W\>\,,
\eeq
where $\<W\>$ is the expectation value of a circular defect of radius $R$. 
In this paper we will consider the change in the defect entropy along the DRG
\beq\label{de}
\delta s(R M)=(1-R\partial_R)\log\frac{\<W_\lambda\>}{\<W_0\>}\,,
\eeq
where 
$\lambda$ is the deformation parameter in (\ref{deltaS}).
This quantity decreases monotonically along the flow \cite{Cuomo:2021rkm, Friedan:2003yc, Casini:2016fgb}. In particular, the change in the $g$-function, $g \equiv \<W_\lambda\>$, between IR and UV fixed points (where the result loses its dependence on $R$), is negative, 
\beq\la{deltag}
\delta g\equiv g_\text{IR}-g_\text{UV}= e^{s_\text{IR}}-e^{s_\text{UV}}=\<W_\lambda\>_\text{IR}-\<W_0\><0\,.
\eeq
It can be thought of as a measure of the number of degrees of freedom that have decoupled along the flow.  
We compute $\delta s$ and $\delta g$ for the $\bar FF$ flow. The calculation is carried out by expanding $\<W_\lambda\>/\<W_0\>$ in powers of $\lambda$ and resuming all orders. 
We interpret our results, which are in agreement with the general $g$-theorem (\ref{deltag}).

One well-known example of a DCFT with ${\mathbb O}_{\bar FF}$ operators is obtained by coupling Chern-Simons theory to fermions in the fundamental representation. There, the conformal defect operator can be a standard Wilson line, and the operator ${\cO}$ is a certain component of the fermion field \cite{Gabai:2022vri, Gabai:2022mya}. In this case, we relate the change in the $g$-function to the change in the two-point function of the displacement operator.

The paper is organized as follows. We begin in section \ref{sec:RGonStraightLine} by analyzing the DRG on a straight line, and determine the fixed points. In section \ref{sec:changeInG} we compute $\delta g$. In section \ref{sec:interpretationsec} we interpret the result by embedding it in a flow that starts from the stable line plus a trivial line. In section \ref{displacementsec} we relate it to the two-point function of the displacement operator in CS-matter theories. Finally, in section \ref{sec:dEalongRG} we analyze the defect entropy along the DRG flow and check its monotonicity.

\section{RG Flow on the Straight Line}
\label{sec:RGonStraightLine}

Our starting point is a straight conformal line defect $\cW$ that ends on a fundamental operator $\cO$ of dimension $\Delta$. We normalize the operator so that its two-point function on the straight defect of length $L$ takes the standard form
\beq\label{line2pf}
\cD_0^\text{line}(L)=\langle \overline{\mathcal{O}}_i(L)\cW[L,0]_{ij}\mathcal{O}_j(0)\rangle\equiv\llangle \overline{\mathcal{O}}(L)\,\mathcal{O}(0)\rrangle
=\frac{1}{L^{2\Delta}}\,,
\eeq
where $i$ and $j$ are color indices.

We assume that $\Delta\in[0,1/2)$ and  
turn on the fundamental anti-fundamental ($\bar FF$) deformation. Due to the planar structure of the large $N$ factorization, the OPE of ${\mathbb O}_{\bar FF}$ with itself consists of ${\mathbb O}_{\bar FF}$ and its descendants only. Hence, no other operator can be generated along the flow. (This is in contrast to the double-trace deformation, where the OPE of two ${\mathbb O}_\text{DT}$ operators includes the operator ${\mathbb O}_\text{DT}^2$ which also becomes relevant for $\Delta<1/4$.)
As a result, the two-point function changes to 
\beq \la{prop}
\cD_\lambda^\text{line}(L)\equiv\llangle \overline{\mathcal{O}}(L)\,\cP\exp\Big[-\lambda M^{1-2\Delta}\int\limits_0^L\dd x\,{\mathbb O}_{\bar FF}(x)\Big]\mathcal{O}(0)\rrangle={1\over L^{2\Delta}}\sum_{n=0}^\infty \({-\lambda\over(ML)^{2\Delta-1}}\)^nJ_n\,,
\eeq
where $\cP$ stands for path ordering, and $J_n$ is the integrated $n$-point function of the operator ${\mathbb O}$ in (\ref{FantiF}). It is given by
\beq\la{integrals}
J_n=\int\limits_{0<\sigma_1<\ldots<\sigma_n<1}{\dd\sigma_1\ldots\dd\sigma_n\over[(1-\sigma_n)(\sigma_n-\sigma_{n-1})\ldots(\sigma_2-\sigma_1)\sigma_1]^{2\Delta}}=\frac{\Gamma\(1-2\Delta\)^{n+1}}{\Gamma\((n+1)(1-2\Delta)\)}\,.
\eeq

The series in (\ref{prop}) sums to 
\beq\la{DofL}
\mathcal{D}_\lambda^\text{line}(L)=\frac{\Gamma(1-2\Delta)}{L^{2\Delta}}E_{1-2\Delta\,,1-2\Delta}\(-\lambda(ML)^{1-2\Delta} \Gamma(1-2\Delta)\)\,,
\eeq
where $E_{\alpha,\beta}(z)$ is the Mittag-Leffler (E) function; see \cite{Tsitouras:2011aa} for a review of its relevant properties. This result can also be obtained by solving the Schwinger-Dyson equation
\beq
\mathcal{D}_\lambda^\text{line}(L)=\mathcal{D}_0^\text{line}(L)-\lambda M^{1-2\Delta}\int\limits_0^L\dd s\,\mathcal{D}_0^\text{line}(L-s)\mathcal{D}_\lambda^\text{line}(s)\,.
\eeq

Because $\Delta<1/2$, all integrals in (\ref{integrals}) are finite and there is no need to introduce a new regularization scale. Hence, no new terms are generated. Here, we are working with the bare operator, and therefore the corresponding $\beta$-function is obtained by simple dimensional analysis from (\ref{deltaS}), $\b_\l=\beta_\text{bare}=-(1-2\D)\l$. It is evident that $\lambda=0$ is an unstable fixed point, with no other fixed point at finite $\lambda$. To analyze the IR limit, we study the asymptotic behavior of the two-point function (\ref{DofL}) at large $ML$. For $\lambda>0$ we have
\beq\la{IRls0}
\cD_{\lambda>0}^\text{line}(L)=\frac{(1-2\D) \sin(2\pi\D)}{\pi(\l M^{1-2\D})^2}\frac{1}{L^{2(1-\Delta)}}\big(1+O\(1/(ML)\)\big)\,.
\eeq
It follows that $\lambda\to\infty$ is a stable fixed point. The dimension of the fundamental operator $\cO$ flows from $\Delta_\cO=\Delta$ in the UV to $\Delta_\cO=1-\Delta$ in the IR. 

Since the flow is super-renormalizable, the only dimensionful scale on which the wave function renormalization factor can depend is $\lambda M^{1-2\Delta}$. This factor is independent of the UV operator's normalization choice. 
It follows that the ratio
\beq
\lim_{ML\to\infty}\(\lambda(ML)^{1-2\Delta}\)^2{\cD_{\lambda>0}^\text{line}(L)\over \cD_0^\text{line}(L)}={1\over\pi}(1-2\Delta)\sin(2\pi\Delta)\,
\eeq
is physical. In section \ref{displacementsec} we will relate this quantity to the change in the two-point function of the displacement operator.

For $\lambda<0$ the two-point function grows exponentially in the IR,
\beq\la{IRg0}
\cD_{\lambda<0}^\text{line}(L)=\frac{M^{2\D}|\l|^{\frac{2\D}{1-2\D}}(\Gamma(1-2\D))^{\frac{1}{1-2\D}}}{1-2\D}\,e^{[-\lambda\Gamma(1-2\Delta)]^\frac{1}{1-2\Delta}ML}\,+\frac{1}{L^{2\D}}O\(1/(ML)\)\,.
\eeq
This deformation of the line therefore destabilizes the IR. It reflects an instability that can affect the theory far from the defect, and we cannot determine its fate on the basis of our assumptions. The same behavior is expected to hold for the double-trace deformation.

In several examples of double-trace deformations that have been studied in \cite{Aharony:2022ntz,Aharony:2023amq} it was found that a small correction to the factorization (which would be of order $1/N$ in our setup) is responsible for restoring conformal invariance in the IR. Assuming that this is the case, in section \ref{sec:interpretationsec}, we will evaluate the corresponding change in the defect entropy.

\subsection*{$\Delta=1/2$} 
The case where $\Delta=1/2$ requires special care because $\mathbb{O}_{\bar FF}$ is classically marginal and the loop integrals in (\ref{integrals}) have logarithmic divergences. We choose to work with a point-splitting regularization, $\sigma_{i+1}-\sigma_i>\epsilon/L$. Working to second order in perturbation theory, we find
\beq\label{exact 2pt D=1}
\cD_\lambda^\text{line}(L)=\frac{1}{L}\left(1-2\l\log\frac{L}{\e}+\l^2(-\frac{\pi^2}{2}+3\log^2\frac{L}{\e})+O(\l^3)\right)\,.
\eeq
To analyze the flow, we plug $\cD$ into the corresponding Callan-Symanzik equation
\beq\label{CS line}
(\e\,\d_\e-\b_\l\d_\l+2\g_\cO)\cD_\lambda^\text{line}(L)=0\,,
\eeq
where $\gamma_\cO$ is the anomalous dimension matrix. From (\ref{CS line}) we find that
\beq\la{betagamma}
\gamma_\cO=-\lambda+O(\lambda^3)\,,\qquad\beta_\lambda=\lambda^2+O(\lambda^3)\,.
\eeq
We see that $\d_\l \b_\l=\l+O(\l^2)$, and so for $\lambda<0$ the $\bar FF$ deformation is marginally relevant while for $\lambda>0$ it is marginally irrelevant. 

\subsection*{Summary}
By comparing the analysis at $\Delta<1/2$ and $\Delta=1/2$, we see that, as $\Delta\to1/2$ from below, the two fixed points collide. As $\Delta$ crosses $1/2$, they interchange with each other. To make this more manifest, we impose the renormalization condition that the two point function of the renormalized boundary operators at $ML=1$ is finite as $\Delta\to1/2$. Using (\ref{prop}) and (\ref{integrals}) we find that this renormalization condition is satisfied by the renormalized operators and coupling
\beq
\lambda\cO=f\cO^\text{ren}
\,,\quad \lambda{\mathbb O}_{\bar FF}=f{\mathbb O}_{\bar FF}^\text{ren}\,,\qquad\text{with}\qquad
f(\lambda)= {\lambda\over1+\lambda/(1-2\Delta)}\,.   
\eeq
In terms of these, the two-point function evaluated at $ML=1$ takes the form
\begin{multline}
\llangle \overline{\mathcal{O}}^\text{ren}(L)\,\cP\,\text{exp}\Big[-f L^{2\Delta-1}\int\limits_0^L\dd x\,{\mathbb O}_{\bar FF}^\text{ren}(x)\Big]\mathcal{O}^\text{ren}(0)\rrangle
\\ =\frac{1}{L^{2\D}}\[1+\sum\limits_{m=1}^\infty \left(\frac{f}{1-2\D}\right)^m\times O(1-2\D)\]\,.
\end{multline}
The corresponding $\beta$-function reads
\beq
\beta_f=f(f-(1-2\Delta))\,.
\eeq
The two fixed points are $f=0$ and $f=1-2\Delta$. The latter (former) is the stable one for $\Delta<1/2$ ($\Delta>1/2$). As $\Delta\to1/2$ the two fixed points collide.

\section{The Change in the $g$-function}
\label{sec:changeInG}

Next, we turn to the computation of the change in the $g$-function between the UV and IR fixed points, (\ref{deltag}). At any point along the flow, the change in the expectation value of the defect takes the form 
\beq\la{ZlZ0}
\delta\<W\>\equiv\<W_\lambda\>-\<W_0\>=\sum_{n=1}^\infty(-\lambda M^{1-2\Delta})^nI_n\,,
\eeq
where $I_n$ is the integrated correlator of $n$ insertions of ${\mathbb O}_{\bar FF}$ on the circle. It is given by a factorized product of two-point functions between neighboring ordered fundamental and anti-fundamental operators. Together, these form a planar connected contraction of ${\mathbb O}_{\bar FF}$, as seen in figure \ref{unregulated}. This integrated correlator is given by
\beq\la{Inintergals}
I_n=R^n\!\!\!\!\!\!\!\!\!\!\!\!\int\limits_{0<\theta_1<\ldots<\theta_n<2\pi}\!\!\!\!\!\!\!\!\!\!\!\!\dd\theta_1\ldots\dd\theta_n\prod_{j=1}^nD_0^\text{arc}(\theta_{j+1}-\theta_j)\,,
\eeq
where $\theta_i$ are angles on the circle, with $\theta_{n+1}\equiv\theta_1+2\pi$. Here, $D_0^\text{arc}(\theta)$ is the two-point function of the operator $\cO$ on an arc of the circle with opening angle $\theta$ and radius $R$ at $\lambda=0$,
\beq\la{D0arc}
\cD_0^\text{arc}(\theta)\equiv\langle \overline{\mathcal{O}}_i(\theta)\cW_\text{arc}[\theta,0]_{ij}\mathcal{O}_j(0)\rangle=\Big[4R^2\sin^2(\theta/2) \Big]^{-\D}\,.
\eeq
It can be obtained by a conformal transformation from the two-point function on a straight line (\ref{line2pf}). 

\begin{figure}[t] 
    \centering
    \includegraphics[width=0.7\textwidth]{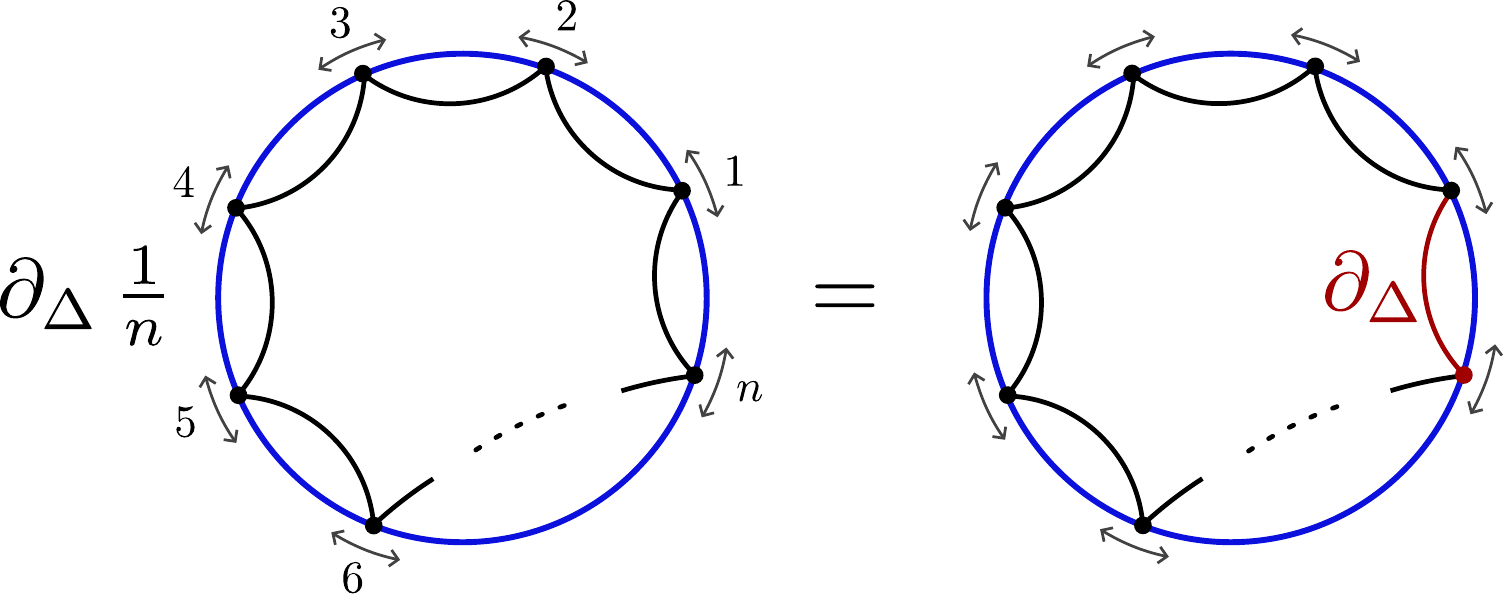} 
   
    \caption{At order $\lambda^n$, the expectation value of the circular loop consists of $n$ ordered contractions between the fundamental and anti-fundamental operators. The insertion points of these operators are integrated around the circle and are labeled by $1,\ldots,n$ in the figure. Upon differentiating by $\D$, we get the derivative of a single contraction, times $n-1$ ordered contractions on an arc. The absence of a combinatorial factor on the right-hand side can be understood as the result of a cancellation between a factor of $1/n$ for the cyclic permutation of the points and a factor of $n$ for the derivative of $n$ contractions.}\label{derivative I_n}
    
\end{figure}

Using the representation of $\delta g$ as the IR value of the sum in (\ref{ZlZ0}) 
we arrive at (see figure \ref{derivative I_n})
\beq\la{SDfordDeltadg}
\d_\Delta\delta g=\lim_{\tilde\epsilon\to0}\(1-R\d_R\)\,2\pi R^2\!\int\limits_{\tilde\epsilon}^{2\pi-\tilde\epsilon}\dd\theta\[\lambda M^{1-2\Delta}\,\cD_\lambda^\text{arc}(2\pi-\theta)_\text{IR}\]\d_\Delta\[\lambda M^{1-2\Delta}\,\cD_0^\text{arc}(\theta)\]\,,
\eeq
where $\tilde\eps$ is a point-splitting regulator on the propagator, and the $\(1-R\d_R\)$ factor removes the local divergence at $\theta \to 0,2\pi$.
Instead, one can simply subtract a local counter term. After doing so, $\d_\Delta\delta g$ is independent of $R$. Here, $\cD_\lambda^\text{arc}(\theta)_\text{IR}$ is the arc two-point function at the IR fixed point. To evaluate it, we note that the wave function renormalization factor in (\ref{IRls0}) is a local property of the operator. For $\D<1/2$, there are no UV divergences on a line segment or an arc, so it is also scheme-independent. Therefore, its value at the IR fixed point is the same on the straight line and on the arc. We conclude that 
\beq\la{DlambdaIR}
\cD_\lambda^\text{arc}(\theta)_\text{IR}=\frac{(1-2\D) \sin(2\pi\D)}{\pi(\l M^{1-2\D})^2}\times\frac{1}{\Big[4R^2\sin^2(\theta/2) \Big]^{1-\D}}\,.
\eeq

By plugging (\ref{D0arc}) and (\ref{DlambdaIR}) into (\ref{SDfordDeltadg}), we arrive at
\beq\la{dDeltaS}
\d_\Delta\delta g
=2\pi(1-2\Delta)\sin(2\pi\Delta)\,.
\eeq
To obtain the change in the $g$-function, we note that at $\Delta=\frac{1}{2}$ the deformation is marginally irrelevant, and therefore $\left.\delta g\right|_{\Delta=\frac{1}{2}}=0$. 
Since $\delta g$ is continuous in $\D$ at this point, we conclude that
\beq\label{deltaSIR}
\delta g=-\frac{\sin(2\pi\Delta)}{\pi}
-(1-2\Delta)\cos(2\pi\Delta)\,.
\eeq

The functions $\delta g$ 
and $\d_\Delta\delta g$ are plotted in figure \ref{deltaSfig}. We note that $\d_\Delta\delta g$ is positive for $0<\D<\frac{1}{2}$ and therefore $\delta g$ increases monotonically in $\D$. Since $\left.\delta g\right|_{\Delta=\frac{1}{2}}=0$, it follows that $g_\text{UV}>g_\text{IR}$ for all $0\leq\D<\frac{1}{2}$, in agreement with the $g$ theorem (\ref{deltag}).\footnote{In CS theory and for $0<\Delta<1/2$, we have $\<W_0\>=\<\cW_\text{unknot}^{{\mathfrak f}=0}\>_{CS}=N\times{\sin(2\pi\Delta)\over\pi(1-2\Delta)}$, which depends on $\Delta$. Using this expression, we find $\d_\Delta\delta s=2\sin ^2(2 \pi  \Delta )\[\sin ^4(2 \pi  \Delta )+(\pi  (1-2 \Delta )+\sin (2 \pi  \Delta ) \cos (2 \pi  \Delta ))^2\]/N$, which is manifestly positive.} 
Notice also that $0\le g_\text{UV}-g_\text{IR}\le1$, with 
$\left.\(g_\text{UV}-g_\text{IR}\)\right|_{\D=0}=1$. This fact is explained in the next section and is to be contrasted with the case of double-trace flow. There, the change in the defect entropy \cite{Diaz:2007an},
\beq\la{deltaSIR ab}
\(\delta s_\text{IR}\)^\text{double-trace}=\frac{\ii}{2 \pi }\[\text{Li}_2\left(e^{2 \ii \pi  \Delta}\right)-\text{Li}_2\left(e^{-2 \ii \pi  \Delta}\right)\]+(1-2\Delta)\log\(2 \sin\pi\D\)\,,
\eeq
is unbounded from below. As a result, the difference $\(g_\text{UV}-g_\text{IR}\)^\text{double-trace}$ is only bounded at order $N$,
\beq
\(g_\text{UV}-g_\text{IR}\)^\text{double-trace}=\<W_0^\text{double-trace}\>(1-e^{\delta s ^\text{double-trace}_\text{IR}})\leq\<W_0^\text{double-trace}\>\,.
\eeq

\begin{figure}[t] 
    \centering
    \includegraphics[width=0.5\textwidth]{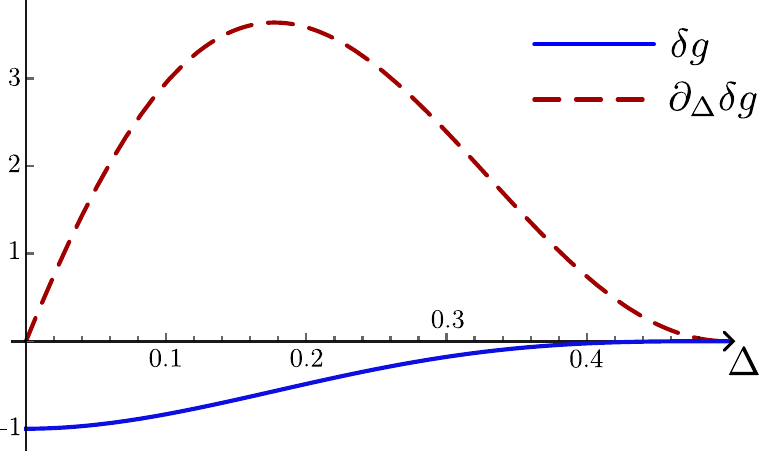} 
   
    \caption{The graph of $\delta g$ and $\d_\D\delta g$ as a function of $\D$. Note that $\d_\D\delta g$ is always positive and, therefore, $\delta g$ increases monotonically.}\la{deltaSfig}
\end{figure}

\section{Interpretation of $g_\text{UV}-g_\text{IR}$} \label{sec:interpretationsec}

We have found that $\delta g$ ranges from $\delta g=0$ for a marginally irrelevant deformation, $2\Delta=1$, to $\delta g=-1$ for $2\Delta=0$. In this subsection, we explain the physics that underlies this result. To this end, 
we first construct the unstable line as the endpoint of a DRG flow from the direct sum of the stable line and the trivial (empty) line. We then use this realization of the unstable line to describe the DRG flow to the stable line that was considered in the previous section. The construction is general and mimics a perturbative procedure as laid out in \cite{Gabai:2022mya}.

Let $\cL_0$ be the one-dimensional Lagrangian of the stable line. For example, in the case where the defect is a Wilson line, $\cL_0$ is proportional to the gauge connection in the direction of the line (which is an $N\times N$ matrix in color space). Using $\cL_0$, the Lagrangian of the direct sum of the stable line and a trivial line can be written as
\beq\la{directsum}
\cL(x)=\begin{pmatrix}
\cL_0(x) & 0\\
0 & 0
\end{pmatrix}\,.
\eeq
It is evident that in the direct sum of lines, there is one more degree of freedom. Consequently, its $g$-function differs by one unit,
\beq
g_{\text{stable}\oplus 1}=\langle\tr\cP\,e^{-\oint\dd x\,\cL(x)}\rangle= \langle\tr\cP\,e^{-\oint\dd x\,\cL_0(x)}\rangle+1=g_\text{stable}+1\,.
\eeq

To construct the unstable line, we deform the line Lagrangian as $\cL\to\cL+\delta\cL$, with
\beq\la{deltaL}
\delta\cL=\alpha\, M^{\Delta}
\begin{pmatrix}
0 & \cO\\
\overline{\cO} & 0
\end{pmatrix}
+
\begin{pmatrix}
0& 0\\
0 & \beta_{c.t.}
\end{pmatrix}
\,,
\eeq
where $M$ is an RG scale, and $\beta_{c.t.}$ 
is a local counter term, that we will fine tune and determine below so that the flow ends at the unstable fixed point. 
This deformation mixes between the stable and trivial lines. When the line is expanded in powers of $\delta\cL$, it decomposes into empty and filled segments, with $\delta\cL$ transitioning between them. At small $\alpha$ the deformation is relevant because $\cO$ has dimension $1-\Delta<1$. Due to the large $N$ factorization property of the correlators of $\mathbb{O}$ on the stable line, no new operator can be generated along the DRG flow.\footnote{As we will see, the operator $\overline\cO\times \cO$ remains irrelevant along this flow.} Assuming that the flow ends at a new conformal line, we can represent the conformal line in the IR as in (\ref{deltaL}) with some specific $\alpha=\alpha_*$. 

To find the new fixed point, we consider the expectation value of two mesonic lines. One is defined before the deformation,
\beq
\cM(L)\equiv\langle
\overline{\cO}(L)\mathcal{P}e^{-\int\limits_0^L\dd x\,\cL_0(x)}\cO(0)\rangle=\langle\overline{\cO}(L)\Big[\mathcal{P}e^{-\int\limits_0^L\dd x\,\cL(x)}\Big]_{11}\cO(0)\rangle
={1\over L^{2(1-\Delta)}}\,,
\eeq
where we have chosen a canonical normalization for $\cO$. The other mesonic line is defined after the line is deformed,
\beq
\widetilde\cM(L)\equiv\langle\Big[\mathcal{P}e^{-\int\limits_0^L\dd x\(\cL(x)+\delta\cL(x)\)}\Big]_{22}\rangle\,.
\eeq
From the structure of $\delta\cL$ in (\ref{deltaL}) it follows that $\widetilde M(L)$ satisfies the differential equation
\beq\la{gen unstab def}
\partial_L\widetilde\cM(L) = \alpha^2M^{2\Delta}\int\limits_0^{L-\e}\dd x\,\cM(L-x)\,\widetilde\cM(x) -\beta_{c.t.}
\widetilde\cM(L-\e)\,,
\eeq
where $\epsilon$ is a point-splitting regulator. Note that here we assume that a cutoff is only needed near $x=L$, and not near $x=0$. That is, point-splitting is only necessary for $\cM$, but not $\widetilde{\cM}$. Consequently, we shall only look for a solution that is consistent with this assumption.
We begin by treating $0<\D<1/2$, excluding the points $0$ and $1/2$. For the divergence near $x=L$ to cancel, we see that
\beq\label{bct}
\beta_{c.t.}={\alpha^2\over1-2\Delta}{M^{2\Delta}\over\epsilon^{1-2\Delta}}\,.
\eeq

At the UV fixed point, the boundary operator $(0,1)^\intercal$ has the lowest dimension, equal to zero. Due to the gap, it cannot mix with $\(\cO,0\)^\intercal$ at small $\alpha$. In fact, from (\ref{gen unstab def}) it follows that at the IR fixed point, $\(\cO,0\)^\intercal_\text{IR}$ is the $SL(2)$ descendant of $\(0,1\)^\intercal_\text{IR}$. In particular, at the IR fixed point $(0,1)^\intercal$ is an $SL(2)$ primary of some dimension $\Delta_\text{IR}$,
\beq\label{unstab conf}
\lim_{(Mx)\to\infty} \widetilde{\cM}(x)\propto \frac{1}{|x|^{2\Delta_\text{IR}}}\(1+O(1/(Mx)\)\,.
\eeq
For this behavior to be consistent with our assumption that the integral in (\ref{gen unstab def}) is finite near $x=0$, we must have $\Delta_\text{IR}<1/2$.

We plug the IR limit (\ref{unstab conf}) into (\ref{gen unstab def}), where the normalization of $\widetilde{M}$ drops out because (\ref{gen unstab def}) is homogeneous. Taking then the limit $\epsilon/L\to0$, we find
\beq\label{unstable eq 0<D<1/2}
0={\Gamma(1-2\Delta_\text{IR})\Gamma(2\Delta-1)\over\Gamma(2\Delta-2\Delta_\text{IR})}(ML)^{2\Delta}-\[{\Delta_\text{IR}\over\Delta}+{2\Delta_\text{IR}\over1-2\Delta}\](M\epsilon)^{2\Delta}+{2\Delta_\text{IR}\over\alpha^2_*}+\text{[subleading]}\,.
\eeq
The local coefficient $\alpha_*$ may depend on the RG and regularization scales, but it cannot depend on $L$. Terms with different powers of $L$ must vanish independently. 
From the cancellation of the $L$ independent terms we conclude that 
\beq\la{IRsol1}
\alpha^2_*M^{2\Delta}=2\Delta(1-2\Delta)\epsilon^{-2\Delta}\,.
\eeq
From the term of order $(ML)^{2\Delta}$, upon restriction to $\D_\text{IR}<1/2$, we can solve for $\D_\text{IR}$,
\beq\la{IRsol}
\Delta_\text{IR}=\Delta\,.
\eeq
Hence, the flow that was triggered by (\ref{deltaL}) has landed at the unstable fixed point of a single line operator.

We can now use (\ref{deltaL}) to represent the flow back to the stable line that we have considered in the previous sections. The $\bar FF$ deformation takes the form
\beq\la{fermionmass}
{\mathbb O}_{\bar FF}=(0,1)^\intercal \times (0,1)=\begin{pmatrix}
0 & 0\\
0 & 1
\end{pmatrix}\,.
\eeq

The unstable line (\ref{deltaL}), (\ref{IRsol1}), (\ref{IRsol}), can be thought of as a condensation of empty segments mixed within the stable line. The relevant $\bar FF$ deformation with positive coefficient leads to an exponential suppression of the empty segments. At the end of the flow that it triggers, the empty segments disappear, and we remain with the stable line only. The direct sum in (\ref{directsum}), (\ref{deltaL}) can equivalently be realized using a worldline fermion, see appendix \ref{worldlinefermion} for details. In this realization, the $\bar FF$ deformation (\ref{fermionmass}) is a mass term for the worldlline fermion. It leads the fermion to decouple in the IR, leaving the stable line.

In summary, we have constructed the unstable line as the fixed point of a DRG from the stable line plus a trivial line. The stable line is then obtained as the fixed point of a second DRG flow that is triggered by (\ref{fermionmass}). Therefore, the $g$-function of the unstable line must lie between that of the stable line and that of the stable plus trivial line,
\beq
g_\text{stable}<g_\text{unstable}<g_\text{stable}+1\,,
\eeq
in agreement with (\ref{deltaSIR}). 

The limit $\D\to0$ requires special care. In this limit, the deformation (\ref{deltaL}) becomes classically marginal and the solution to (\ref{gen unstab def}) needs to be reconsidered. 
We find that for $\D=0$ the integral in (\ref{gen unstab def}) has a logarithmic divergence, which can only be eliminated by setting $\a_*^2(\D=0)=0$. It follows that $\delta\cL$ is marginally irrelevant and $\D_\text{IR}=0=\D$. 
We conclude that in this limit $g_\text{unstable}\to g_\text{stable}+1$, in agreement with (\ref{deltaSIR}).

\subsection{$\delta g$ for the $\lambda<0$ Flow}

Realizing the unstable fixed point using (\ref{deltaL}) is also useful for understanding the flow with $\lambda<0$, (\ref{IRg0}). For negative $\lambda$ the $\bar FF$ deformation leads to an exponential enhancement of the empty segments. At the end of this flow, the empty segments dominate and we remain with the empty line only. If we assume that a small correction (of order $1/N$) is responsible for restoring conformality in the IR, then the change in the defect entropy is $\delta g=1-g_\text{unstable}+O(1/N)$, which can be large due to the large value of $g_\text{unstable}$.

\section{Relation to the Two-Point Function of the Displacement Operator in CS-matter Theories}\la{displacementsec}

One characteristic of any conformal line operator is the two-point function of the displacement operator on the circle, which we denote by $\gamma$. The $\bar FF$ flow that we have studied here is realized in CS-matter theory. In that case, the two-point function of the displacement operator was computed in \cite{Gabai:2022mya} and is given by\footnote{Similar looking results were obtained in \cite{Giombi:2013yva} for the change in the three-sphere free energy of large N CFTs deformed by double-trace operators. We thank Igor Klebanov for pointing this out to us.}
\beq\la{displacement2pf}
\gamma={\Lambda(\Delta)\over\<W_0\>}\,,\qquad \Lambda(\Delta)=-\frac{1}{2 \pi }(2\Delta -1) (2 \Delta -2) (2 \Delta -3) \sin(2 \pi \Delta)\,.
\eeq
Here, $\Lambda$ is the two-point function on the mesonic line \cite{Gabai:2022mya}, and $\<W_0\>\propto N$ is the expectation value of the unknot Wilson loop in CS theory. In this section, we relate the changes between the two fixed points of the defect entropy, the two-point function of the displacement operator, and the wave function renormalization factor.

\subsubsection*{The Change in the Displacement Operator} 
The DRG is triggered in the UV by deforming the line action with the $\bar FF$ operator (\ref{FantiF}). The corresponding UV change in the displacement operator that follows is
\beq\la{deltaDpm}
\delta{\mathbb D}_\pm=-\lambda M^{1-2\Delta}\[\delta_\pm\cO\times\overline\cO+\cO\times\delta_\pm\overline\cO\]\,,
\eeq
where $\delta_\pm\cO$ stands for the path derivative of $\cO$, see \cite{Gabai:2022vri,Gabai:2022mya,Gabai:2023lax} for details. Consider first the deformation of ${\mathbb D}_+$. We have $\delta_+\cO=\epsilon^{1-2\Delta}\cO_+$, where $\epsilon$ is a UV cutoff renormalization scale, and $\cO_+$ is an operator of dimension $\Delta_+=2-\Delta$. 
This term, with a positive power of $\epsilon$, vanishes as we remove the cutoff $\epsilon\to0$. In the second term, $\delta_+\overline\cO=\overline\cO_+$ is an operator of dimension $\overline\Delta_+=\Delta+1$. In the IR, $M$ is large and the second term in (\ref{deltaDpm}) dominates the displacement operator,
\beq
\({\mathbb D}_+\)_\text{IR}=-\lambda M^{1-2\Delta}\cO\times\overline\cO_+\,.
\eeq
Similarly,
\beq
\({\mathbb D}_-\)_\text{IR}=-\lambda M^{1-2\Delta}\cO_-\times\overline\cO\,.
\eeq

The operators $\overline\cO_+$ and $\cO_-$ carry a transverse spin that is different from that of $\cO$ and $\overline\cO$ respectively. As a result, their two-point function is blind to the $\bar FF$ deformation and does not change along the flow. In the UV, this two-point function obeys (see \cite{Gabai:2022vri,Gabai:2022mya,Gabai:2023lax})
\beq
\llangle \overline\cO_+(L)\,\cO_-(0)\rrangle_\text{UV}={1\over2}\d_L^2\llangle \overline\cO(L)\,\cO(0)\rrangle_\text{UV}\,.
\eeq
The operators $\cO$ and $\overline\cO$ are normalized as in (\ref{line2pf}). Therefore,
\beq
\llangle \overline\cO_+(L)\,\cO_-(0)\rrangle_\text{IR}=\llangle \overline\cO_+(L)\,\cO_-(0)\rrangle_\text{UV}=
\frac{\Delta(2\Delta+1)}{L^{2\Delta+2}}\,.
\eeq
On the other hand, the two-point function of $\cO$ and $\overline\cO$ changes along the flow. In the IR it is given in (\ref{IRls0}), which we repeat here for convenience,
\beq\la{wfrf}
\llangle \overline\cO(L)\,\cO(0)\rrangle_\text{IR}
=\frac{(1-2\D) \sin(2\pi\D)}{\pi(\l M^{1-2\D})^2}\frac{1}{L^{2(1-\Delta)}}\,.
\eeq

It follows that the two-point function of the displacement operator in the IR is equal to\footnote{In CS coupled to fundamental bosons, we further have $\Lambda_\text{IR}-\Lambda_\text{UV}={3\over\pi^2}\lambda_b\d_{\lambda_b}\delta g$, where $\l_b$ is the 't Hooft coupling. A similar looking relation holds for the circular 1/2 BPS Wilson loop in $\mathcal{N}=4$ SYM theory and is related to supersymmetry \cite{Correa:2012at}.}
\beq\la{deltaLambda}
\Lambda(\Delta)_\text{IR}=\[\Delta(2\Delta+1)\]\times\[{1\over\pi}(1-2\D) \sin(2\pi\D)\]=\Lambda(1-\Delta)\,,
\eeq
in agreement with the expression (\ref{displacement2pf}) and a flow from $\Delta$ and $1-\Delta$.

\section{The Defect Entropy Along the RG flow}
\label{sec:dEalongRG}

We now turn to the computation of the change in the defect entropy (\ref{de}) along the $\bar FF$ DRG flow. We will then use this result to check the monotonicity of the defect entropy along the flow explicitly.

The change in the expectation value of the defect $\delta\<W\>$ is of order one, while $\<W_0\>$ is of order $N$. Therefore,  
\beq\la{deltasofR}
\delta s=(1-R\partial_R)\log\frac{\<W_\lambda\>}{\<W_0\>}= (1-R\partial_R){\delta\<W\>\over\<W_0\>}\(1+O(1/N)\)\,.
\eeq
The expansion of $\delta\<W\>$ in powers of $\lambda$ is given in (\ref{ZlZ0}).  
To evaluate $I_n$ in (\ref{Inintergals}), it will be useful to first relax the condition that $\theta_{n+1}=\theta_1+2\pi$, and let $\theta_2,\ldots,\theta_{n+1}$ run on the universal cover of the circle,
\beq\la{tIn}
\tilde I_n(a)\equiv R^{n+1}
\int\limits_0^{2\pi} \dd \th_1\!\!\!\!\!\!\!\!\!\!\!\!\!\!\!
\int\limits_{\qquad\theta_1<\th_2<\ldots<\theta_{n+1}}\!\!\!\!\!\!\!\!\!\!\!\!\!\!\!\!\dd\theta_2\ldots\dd\theta_{n+1}\,\prod_{j=1}^nD_0^\text{arc}(\theta_{j+1}-\theta_j)\,e^{\ii a(\theta_{j+1}-\theta_j)}\,.
\eeq
Here, we have dressed $D_0^\text{arc}(\theta_{j+1}-\theta_j)$ by $e^{\ii a(\theta_{j+1}-\theta_j)}$. These factors telescope to $e^{\ii a(\theta_{n+1}-\theta_1)}$. We then use this phase factor to impose the condition $\theta_{n+1}\equiv\theta_1+2\pi$ as
\beq\la{IatoI}
I_n=\frac{1}{n}\int\limits_{-\infty}^{\infty}{\dd a\over2\pi R}\,e^{-\ii a\cdot 2\pi}\tilde I_n(a)\,,
\eeq
where the combinatorial factor of $1/n$ accounts for cyclic permutations of the $n$ points on the circle (there was no such combinatorial factor in (\ref{Inintergals}) because the first point was specified).

\subsection{Divergences and Regularization}

The loop integrals in (\ref{Inintergals}) have potential divergences from the integration regions where some of the insertion points collide. Consider first the region where only a subset of $m<n$ consecutive points collide. 
In this region, the integral behaves in the same way as the line integral, $J_n$ in (\ref{integrals}), behaves in the region where $m$ insertion points collide.
Because $J_n$ is finite, this region does not cause a divergence. To see this more explicitly, we denote by $\rho$ the common relative distance between these insertion points. Near $\rho=0$, the integral behaves as $\int\rho^{m-2}\dd\rho/\rho^{2(m-1)\Delta}$. Here, the numerator comes from the measure after factoring out the center of mass of these $m$ points. The denominator comes from the $m-1$ contractions between the $m$ insertion points. This integral is finite for $\Delta<1/2$. Therefore, the only potential divergence comes from the integration region where all $n$ points collide. In this region, all $m=n$ contractions are singular (instead of only $m-1$ for $m<n$). Consequently, the integral behaves as $\int\rho^{n-2}\dd\rho/\rho^{2n\Delta}\propto\epsilon^{n(1-2\Delta)}/\epsilon$, where $\epsilon$ is a UV cutoff on $\rho$. We conclude the following:
\begin{itemize}
\item For any $\Delta<1/2$, only finitely many $I_n$'s are divergent.
\item The divergences are linear in $R$ and can be removed by a local counter term. They are also removed by the factor $(1-R\d_R)$ in the defect entropy (\ref{de}).
\item To regulate these divergences, it suffices to regulate one of the $n$ insertions.
\end{itemize}
In the following, we choose to regulate the $I_n$ integrals by point-splitting the insertion at $\theta_1$ as
\beq
{\mathbb O}_{\bar FF}(\theta_1)=\overline{\cO}(\theta_1) \times \cO(\theta_1)\quad\rightarrow\quad\overline{\cO}(\theta_1) \times \cO(\theta_1+\tilde\epsilon)\,,
\eeq
where $\tilde\epsilon=\epsilon/R$ is the regulator. This is achieved by modifying (\ref{IatoI}) to
\beq\la{Ireg}
I_n^\text{reg}\equiv\frac{1}{n}\int\limits_{-\infty}^{\infty}{\dd a\over2\pi R}\,e^{-\ii a\cdot (2\pi-\tilde\epsilon)}\tilde I_n(a)\,.
\eeq

\subsection{Evaluation of $\delta s$}

The integrals in $\tilde I_n$, (\ref{tIn}), 
are ordered convolutions. They are therefore diagonalized by a Laplace transform
\beq\label{lap}
\cL(s)\equiv R^{2\D}\int\limits_{0}^{\infty} \dd\theta\, D_0^\text{arc}(\theta) e^{-s\theta}=\frac{\pi \Gamma(1-2\Delta)\,\text{csch}(\pi s)}{\Gamma\left(1-\D-\ii s\right)\Gamma\left(1-\D+ \ii s\right)}\,.
\eeq
The inverse transform reads
\beq
D_0^\text{arc}(\theta)=\frac{1}{2\pi \ii\,R^{2\D}} \int\limits_{\cal C}\! \dd s\, \cL(s) e^{s\theta}\,,
\eeq
where the path ${\cal C}$ is parallel to the imaginary axis at some positive real value, and is oriented towards the positive imaginary $s$ axis.
Here, for convenience, we have factored out the dependence on $R$ so that $\cL(s)$ is dimensionless.

Using the convolution relation
\beq
\int\limits_{\alpha}^{\beta} \dd \theta\, D_0^\text{arc}(\beta-\theta)D_0^\text{arc}(\theta-\alpha)=\frac{1}{2\pi\ii R^{4\D}}\int\limits_{\cal C} \dd s \,\cL^2(s)\, e^{s(\beta-\alpha)}\,,
\eeq
the regulated integral (\ref{Ireg}) takes the form
\beq\la{In1}
    I_n^\text{reg}=-{\ii\over n}\,R^{(1-2\D)n}\int\limits_{\cal C}\dd s\,\cL^n(s)\,e^{(2\pi-\tilde\epsilon) s}\,.
\eeq
In this fashion, the regulated partition function resums to
\beq\label{deltaW}
\delta\<W\>^\text{reg}=\ii\int\limits_{\cal C}\dd s\,e^{(2\pi-\tilde\epsilon) s} \log\[1+\lambda(MR)^{1-2\Delta}\cL(s)\]+ O(1/N)\,.
\eeq
The renormalized operator is obtained from (\ref{deltaW}) by subtracting the local divergent contributions. For $\Delta<1/4$ the only divergent contribution is the term linear in $\lambda$, which is due to the self-contractions of the two operators in ${\mathbb O}_{\bar FF}$, (\ref{FantiF}). After subtracting it, one can take $\tilde\epsilon\to0$. Other values of $\Delta$ can be obtained by analytically continuing the result in $\Delta$ or, equivalently, by subtracting additional divergent terms. These local divergent terms drop out from the change in the defect entropy (\ref{deltasofR}).

\subsection{IR Limit}
We can now check that the change in the $g$-function (\ref{deltaSIR}) is reproduced from (\ref{deltaW}) upon taking the IR limit, $MR\to \infty$. Recalling that $\Delta<\frac{1}{2}$ and $\lambda>0$, we see that in this limit 
the one inside the logarithm in (\ref{deltaW}) is negligible. Using 
\beq\la{zeroint}
\int\limits_{\cal C}\dd s\, e^{(2\pi-\tilde\epsilon) s} \propto \delta(2\pi-\tilde\e)=0\,,
\eeq
all the $s$-independent terms from (\ref{lap}) drop out and we remain with
\beq
\delta\<W\>^\text{reg}_\text{IR}=-\ii\int\limits_{\cal C}\dd s\,e^{(2\pi-\tilde\epsilon) s}\log\[\sinh(\pi s)\Gamma\big(1-\D-\ii s\big){\Gamma\big(1-\D+ \ii s\big)}\]\,.
\eeq

To evaluate this integral, we take a derivative with respect to $\Delta$
\beq
\d_\Delta\delta\<W\>^\text{reg}_\text{IR}=\ii\int\limits_{\cC}\dd s\,e^{(2\pi-\tilde\epsilon) s}\[\psi\big(1-\D- \ii s\big)+\psi\big(1-\D+ \ii s\big)\]\,,
\eeq
and then use the sum representation of the digamma function. After dropping the $s$-independent terms using (\ref{zeroint}) and interchanging the order of summation and integration, we remain with
\beq\label{sum_inv_lap} 
\d_\Delta\delta\<W\>^\text{reg}_\text{IR}=-\ii\sum_{n=1}^\infty\int\limits_{\cC}\dd s\,e^{(2\pi-\tilde\epsilon) s} 
\frac{2(n-\D)}{(n-\D)^2+s^2}\,.
\eeq
By closing the contour towards the negative real axis and picking the two conjugate poles these integrals evaluate to
\beq
\int\limits_{\cC}\dd s\,e^{(2\pi-\tilde\e) s} 
\frac{2(n-\D)}{(n-\D)^2+s^2}
=2\pi\left(e^{\ii(n-\D)(2\pi-\tilde\epsilon)}-e^{-\ii(n-\D)(2\pi-\tilde\epsilon)}\right)\,.
\eeq
The sum in (\ref{sum_inv_lap}) then becomes a geometric series of the form $\sum_{n=1}^\infty e^{\pm\ii n\varphi}=\frac{e^{\pm\ii\varphi}}{1-e^{\pm\ii\varphi}}$, with $\varphi=2\pi-\tilde\epsilon$,
\begin{align}
\d_\Delta\delta\<W\>^\text{reg}_\text{IR}=&-2\pi\ii\Big[ \frac{e^{\ii\varphi}e^{-\ii\D\varphi}}{1-e^{\ii\varphi}}-\frac{e^{-\ii\varphi}e^{\ii\D\varphi}}{1-e^{-\ii\varphi}}\Big]\\
=&
2\pi\Big[-\frac{2\cos(2\pi\Delta)}{\tilde\epsilon}+(1-2\Delta)\sin(2\pi\Delta)+O(\tilde\epsilon)\Big] \,,\nn
\end{align}
where in the second step we have expanded the result at small $\tilde\epsilon$. The $1/\tilde\epsilon=R/\epsilon$ divergence is linear in $R$ and is removed by a local counter term. 
We thus find
\beq\la{dDeltaS2}
\d_\Delta\(g_\text{IR}-g_\text{UV}\)
=2\pi(1-2\Delta)\sin(2\pi\Delta)\,,
\eeq
in agreement with (\ref{dDeltaS}).

\subsection{Monotonicity of the Defect Entropy}

The monotonicity of the defect entropy has been proven in \cite{Cuomo:2021rkm} by relating its derivative with respect to the RG scale to the two-point function of the defect stress tensor $T_D$,
\beq\label{stress entropy raw}
M\d_M \delta s= -2\pi R^2\int\limits_0^{2\pi}\dd\th\, \langle T_D(0)T_D(\th)\rangle (1-\cos\th)\,.
\eeq 
In this section, we explicitly verify this relation for the $\bar FF$ flow. This serves as a consistency check and also as a more direct proof of the monotonicity in $MR$ of our result for $\delta s$, (\ref{deltaW}).

For the purpose of checking this relation, we take $0<\Delta<1/4$ and subtract the only divergent contribution in $\delta\<W\>^\text{reg}$ in this range, which is $I_1$ in (\ref{ZlZ0}), 
\beq\la{deltaWreg}
\delta\<W\>^\text{ren}=\sum_{n=2}^\infty(-\lambda M^{1-2\Delta})^nI_n\,,\qquad\text{for}\quad0<\Delta<1/4\,.
\eeq
Other values of $\D\in[0,1/2]$ are obtained from $\Delta\in(0,1/4)$ by analytic continuation. When working with $\delta\<W\>^\text{ren}$ we can set $\tilde\epsilon=0$ because all $I_n$'s are finite.

The defect stress tensor along a DRG flow is given by $T_D=\sum_i \b_i\cO_i$, where the sum is over all defect primary operators $\cO_i$ that participate in the flow, and $\b_i$ are the beta functions of their couplings. In our case, due to the large $N$ factorization, the only primary operator that appears in the two-point function of the $\bar FF$ operator (\ref{FantiF}) 
is the $\bar FF$ operator itself. 
Using the result for the beta function found previously, we conclude that
\beq
T_D(\theta)=-(1-2\D)\l \,M^{1-2\D}\,\mathbb{O}_{\bar FF}(\theta)\,.
\eeq
Correspondingly, (\ref{stress entropy raw}) takes the form
\begin{multline}\label{stress entropy}
M\d_M (1-R\d_R)\delta\<W\>^\text{ren}
\\= -2\pi\[(1-2\D)\l \,(RM)^{1-2\D}\]^2\,R^{4\D}\,\int\limits_0^{2\pi}\dd\th\,\cD_\lambda^\text{arc}(2\pi-\th)\cD_\lambda^\text{arc}(\th) (1-\cos\th)\,,
\end{multline} 
where we have canceled the normalization factor of $\<W_0\>$ on both sides. Here, $\cD_\lambda^\text{arc}(\th)$ is the arc two-point function. It can be expanded as was done in (\ref{prop}), on the circle instead of the line,  
\beq\la{IatoI2}
\cD_\lambda^\text{arc}(\th)=\sum_{n=0}^\infty(-\lambda M^{1-2\Delta})^n\,{\mathcal I}_n(\theta)\,,\qquad\text{with}\qquad
{\mathcal I}_n=\int\limits_{-\infty}^{\infty}{\dd a\over(2\pi R)^2}\,e^{-\ii a\, \theta}\tilde I_{n+1}(a)\,.
\eeq
Its Laplace transform can be evaluated as a convolution as was done for $I_n$ (\ref{In1}), and takes the form 
\beq\label{LD}
\cL_\cD(s)=R^{2\D}\int\limits_{0}^{\infty} \dd\theta\, \cD_\lambda^\text{arc}(\theta) e^{-s\theta}=\sum_{n=0}^\infty\(-\l (MR)^{1-2\D}\)^n\cL^{n+1}(s)\,.
\eeq

On the right-hand side of (\ref{stress entropy}) we have the convolution of two functions on the circle. This is given by
\begin{multline}\label{fullRhs}
R^{4\D}\int\limits_0^{2\pi}\, \dd \th\, \cD_\lambda^\text{arc}(2\pi-\th)\,\cD_\lambda^\text{arc}\,(\th) (1-\cos\th)\\=\int\limits_{\cC}\,\frac{\dd s}{2\pi \ii}e^{2\pi s}\,\cL_\cD(s)\left[\cL_\cD(s)-\frac{\cL_\cD(s-\ii)+\cL_\cD(s+\ii)}{2}\right]\,.
\end{multline}
The term of order $\[-\l (MR)^{1-2\D}\]^{n-2}$ in (\ref{fullRhs}) can then be expressed as an inverse Laplace transform evaluated at $2\pi$,
\beq\la{LTT}
\Big[R^{4\D}\int\limits_0^{2\pi}\, \dd \th\, \cD_\lambda^\text{arc}(2\pi-\th)\,\cD_\lambda^\text{arc}\,(\th) (1-\cos\th)\Big]_{n-2}=\int\limits_{\cC}\,\frac{\dd s}{2\pi \ii}e^{2\pi s} f_n(s)\,,
\eeq
of the function
\begin{align}\label{LTTf}
f_n(s)=&
\sum_{m=0}^{n-2}\cL^{m+1}(s)\left[\cL^{n-m-1}(s)-\frac{\cL^{n-m-1}(s-\ii)+\cL^{n-m-1}(s+\ii)}{2}\right]\\
=&(n-1)\cL^{n}(s)-\frac{1}{2}\frac{\cL^{n}(s)\cL(s-\ii)-\cL(s)\cL^{n}(s-\ii)}{\cL(s)-\cL(s-\ii)} -\frac{1}{2}\frac{\cL^{n}(s)\cL(s+\ii)-\cL(s)\cL^{n}(s+\ii)}{\cL(s)-\cL(s+\ii)}\,.\nn
\end{align}

The integral on the right hand side of (\ref{LTT}) is convergent for every $\D\in(0,1/2)$. For $\D\in(0,1/4)$ it converges for each of the five terms in (\ref{LTTf}) separately. This allows us to shift the integration variable in the inverse Laplace transform of the terms with $\cL^n(s\pm\ii)$ by one imaginary unit, such that (\ref{LTT}) holds with
\beq
f_n(s)\to\tilde f_n(s)=\cL^n(s)\Big[(n-1)-\frac{\cL(s-\ii)}{\cL(s)-\cL(s-\ii)}-\frac{\cL(s+\ii)}{\cL(s)-\cL(s+\ii)}\Big]\,.
\eeq
Finally, we substitute in the specific form of $\cL(s)$ from \eqref{lap} and obtain
\beq\label{RHSlap}
 \Big[-2\pi R^2\<W_0\>\int\limits_0^{2\pi}\, \dd \th \langle T_D(0)T_D(\th)\rangle (1-\cos\th)\Big]_n=\ii(1-2\D)(n(1-2\D)-1)\int\limits_{\cC}\,\dd s\,e^{2\pi s}\cL^{n}(s)\,.
\eeq

For the left hand side of (\ref{stress entropy}), by substituting (\ref{In1}) into (\ref{deltaWreg}) we can express the term of order $\[-\l (MR)^{1-2\D}\]^n$ as
\beq\la{inversLlhs}
\[M\d_M (1-R\d_R)\delta\<W\>^\text{ren}\]_n=\ii(1-2\D)(n(1-2\D)-1)\int\limits_{\cC}\dd s\,e^{2\pi s}
\cL^{n}(s)\,,
\eeq
which concurs exactly with \eqref{RHSlap}. This concludes the verification of (\ref{stress entropy raw}) for our case of $\bar FF$ flow.
We note that an analogous derivation also applies to the monotonicity of the defect entropy under a double-trace deformation (\ref{dtdeformation}).

\acknowledgments
We thank Z. Komargodski, I. Klebanov, and S. Yankielowicz, for valuable discussions and comments on the manuscript. AS thanks the Simons Center for Geometry and Physics for its hospitality. IN thanks Z. Bajnok and E{\"o}tv{\"o}s Lor{\'a}nd University for their warm hospitality. AS is supported by the Israel Science Foundation, grant number 1197/20. DlZ is supported in part by the Royal Society University Research Fellowships grant URF/R1/221310 ``Bootstrapping Quantum Gravity''.

\appendix

\section{Worldline Fermion Realization}\la{worldlinefermion}

In section \ref{sec:interpretationsec} we constructed the unstable fixed point using a two-dimensional linear space on the line. One way of realizing such a space is by thinking of it as the Hilbert space of a worldline fermion. In this appendix, we elaborate on this realization.

Consider two worldline fermions $\chi_i(\tau)$, $i=1,2$, where $\tau$ is some parametrization of the line. The worldline Lagrangian reads $\mathcal{L}_\chi=\frac{1}{2}\chi_i\d_\tau\chi^i$. The corresponding equation of motion states that $\chi^i$ is constant. Upon quantization, the canonical anti-commutation relations are $\{\chi^i(\tau),\chi^j(\tau)\}=2\delta^{ij}$. The irreducible representation of these relations are constant $2\times2$ matrices, which we choose to be the Pauli matrices $\chi^i(\tau)=\sigma^i$.

Using this worldline representation, the direct sum of the stable line and a trivial line (\ref{directsum}) corresponds to the worldline Lagrangian $\cL=\cL_\chi+\chi^+\chi^-\,\cL_0(x(\tau))$, where $x(\tau)$ parametrizes the line defect and $\chi^\pm=(\chi^1\pm\ii\,\chi^2)/2$. Similarly, the deformation of the line (\ref{deltaL}) takes the form $\delta\cL=\alpha M^\Delta\(\chi^+\mathcal{O}+\chi^-\overline{\cO}\)+\beta_\text{c.t.}\chi^-\chi^+$. Finally, the DRG flow from the unstable to the stable fixed point is triggered by the operator $\chi^-\chi^+$, which is a mass term on the worldline. 

In summary, the action of the worldline fermion is quadratic. Upon integrating out $\chi^i$ we end up with the $2\times2$ matrix structure of section \ref{sec:interpretationsec}.

\bibliography{bib}
\bibliographystyle{JHEP}

\end{document}